\documentclass[sigconf,nonacm]{acmart}

\usepackage{multirow}
\usepackage{balance}
\usepackage{mathtools}
\usepackage{bm}

\begin{document}

\title[Identifying Offline Metrics that Predict Online Impact: A Pragmatic Strategy for Real-World Recommender Systems]{Identifying Offline Metrics that Predict Online Impact: A Pragmatic Strategy for Real-World Recommender Systems}

\author{Timo Wilm}
\orcid{0009-0000-3380-7992}
\email{timo.wilm@otto.de}
\affiliation{%
  \institution{OTTO (GmbH \& Co. KGaA)}
  \city{Hamburg}
  \country{Germany}
}
\author{Philipp Normann}
\orcid{0009-0009-5796-2992}
\email{philipp.normann@tuwien.ac.at}
\affiliation{%
  \institution{TU Wien}
  \city{Vienna}
  \country{Austria}
  \vspace{0.6cm}
}

\acmArticleType{Research}

\acmCodeLink{https://github.com/otto-de/MultiTRON}

\keywords{offline–online evaluation, offline evaluation, online evaluation, pareto front, session-based recommender systems}

  \ccsdesc[500]{Information systems~Evaluation of retrieval results}
  \ccsdesc[500]{Information systems~Recommender systems}
  \ccsdesc[500]{Computing methodologies~Multi-task learning}
  \ccsdesc[500]{Computing methodologies~Neural networks}

\begin{abstract}
  A critical challenge in recommender systems is to establish reliable relationships between offline and online metrics that predict real-world performance. Motivated by recent advances in Pareto front approximation, we introduce a pragmatic strategy for identifying offline metrics that align with online impact. A key advantage of this approach is its ability to simultaneously serve multiple test groups, each with distinct offline performance metrics, in an online experiment controlled by a single model. The method is model-agnostic for systems with a neural network backbone, enabling broad applicability across architectures and domains. We validate the strategy through a large-scale online experiment in the field of session-based recommender systems on the OTTO e-commerce platform. The online experiment identifies significant alignments between offline metrics and real-word click-through rate, post-click conversion rate and units sold. Our strategy provides industry practitioners with a valuable tool for understanding offline-to-online metric relationships and making informed, data-driven decisions.
\end{abstract}

\maketitle
\vspace{0.25cm}
\makeatletter{\renewcommand*{\@makefnmark}{}
\footnotetext{© Timo Wilm, and Philipp Normann 2025.
This is the author’s version of "Identifying Offline Metrics that Predict Online Impact: A Pragmatic Strategy for Real-World Recommender Systems". It is posted here for your personal use. Not
for redistribution. The definitive version of record was accepted for publication in the
19th ACM Conference on Recommender Systems (RecSys 2025). The final published version will be available at the ACM Digital Library: \\ ACM ISBN: 979-8-4007-1364-4/2025/09  \\ https://doi.org/10.1145/3705328.3748111\makeatother}

\section{Introduction}
Large-scale e-commerce platforms, such as OTTO, face the challenge of optimizing recommender systems to meet diverse business objectives. These systems typically optimize offline metrics to enhance online Key Performance Indicators (KPIs) \cite{jannach_measuring_2019,wilm_scaling_2023,mei_lightweight_2022}. A gap remains between offline metrics and real-world KPIs \cite{rossetti_contrasting_2016,najmani_offline_2022,garcin_offline_2014}. Bridging it is essential to ensure that offline optimizations result in online improvements.

For industry practitioners, an efficient and reliable tool to accurately identify offline metrics aligned with online KPIs is critical for making informed decisions and accelerating optimization cycles. This work presents a pragmatic strategy leveraging recent advances in Pareto front approximation \cite{wilm_pareto_2024,chen_gradient-based_2025,zhang_pmgda_2025} to address this gap. Our strategy enables the simultaneous splitting of traffic into multiple test groups, each with distinct offline metrics, while serving all groups through a single scalable model. This facilitates systematic measurement of offline-to-online metric relationships at scale.

\vspace{0.0cm}
\section{Related Work}
Understanding the relationship between offline metrics and their ability to predict online KPIs has been a central focus in recommender systems research and among industry practitioners. While the academic community often relies on offline benchmarks \cite{canamares_offline_2020,castells_offline_2022} due to limited access to online A/B testing infrastructure, such metrics occasionally fail to generalize to real-world KPIs \cite{garcin_offline_2014,rossetti_contrasting_2016,edizel_towards_2024}.

In contrast, practitioners in industry can conduct online evaluations, but these are typically costly, time-consuming, and operationally complex, particularly at scale. This disconnect has driven a growing body of research aimed at bridging the gap between offline and online evaluation \cite{elahi_online_2024,najmani_offline_2022,peska_off-line_2020,wang_how_2023}.

To circumvent the need for online experiments, off-policy and counterfactual evaluation methods have been proposed. However, these approaches face inherent bias-variance trade-offs that limit their scalability \cite{jeunen_-ope_2024,swaminathan_off-policy_2017,mcinerney_counterfactual_2020,narita_debiased_2021}. Complementary research explores simulation environments that model user behavior more realistically than static user logs \cite{krauth_offline_2020,aouali_offline_2022}. Other research directions have focused on improving offline evaluation through sampling strategies, penalizing popularity bias or incorporating temporal dynamics from user transaction histories to better align offline metrics with online KPIs \cite{kasalicky_bridging_2023,carraro_debiased_2020,carraro_sampling_2022}.

An empirical study on a small-scale e-commerce platform has examined the correlation between offline metrics and online KPIs by deploying multiple models in parallel \cite{peska_off-line_2020}. This method is often infeasible for large-scale systems due to the high cost of training, deploying, and maintaining several model variants simultaneously.

Recently, Pareto front approximation techniques have been scaled to deep recommender systems, enabling effective modeling of trade-offs among competing objectives during inference using a single model at industry scale \cite{wilm_pareto_2024,zhang_pmgda_2025}. Accessing the entire Pareto front at inference time offers a scalable strategy to identify offline metrics that reliably predict online KPIs, making this a promising approach to bridge the offline-to-online evaluation gap.

\bigbreak
\section{Contributions}
We present a pragmatic strategy for identifying offline metrics that predict online KPIs in large-scale recommender systems, enabling more efficient, data-driven decision-making for industry practitioners. Our main contributions are as follows:
\begin{itemize}
    \item A strategy based on Pareto front approximation that identifies offline metrics capable of predicting online outcomes. The approach is model‑agnostic for neural backbones.
    \item We demonstrate the effectiveness of our strategy through an online experiment in the field of session-based recommender systems on the OTTO e-commerce platform.
    \item We define a novel offline metric, \emph{order density (OD)}, which estimates the post-click conversion rate. The product metric \emph{Recall@20 $\cdot$ OD@20} serves as an offline proxy for units sold, while Recall@20 estimates the click-through rate.
    
    \item Along the efficient frontier, sacrificing \emph{OD@20} to increase \emph{Recall@20} proves to be a more efficient strategy for driving units sold on the OTTO platform.
    
\end{itemize}
To extend the applicability of our strategy to single-objective systems, we introduce an auxiliary \emph{distortion loss}, an artificial objective that forms the second axis of the Pareto front. For future research and applications, \emph{OD@20}, \emph{Recall@20 $\cdot$ OD@20}, and the \emph{distortion loss} have been incorporated into the MultiTRON\footnote[1]{\url{https://github.com/otto-de/MultiTRON}} source code.

\begin{figure*}
  \centering
  \includegraphics[scale=0.56]{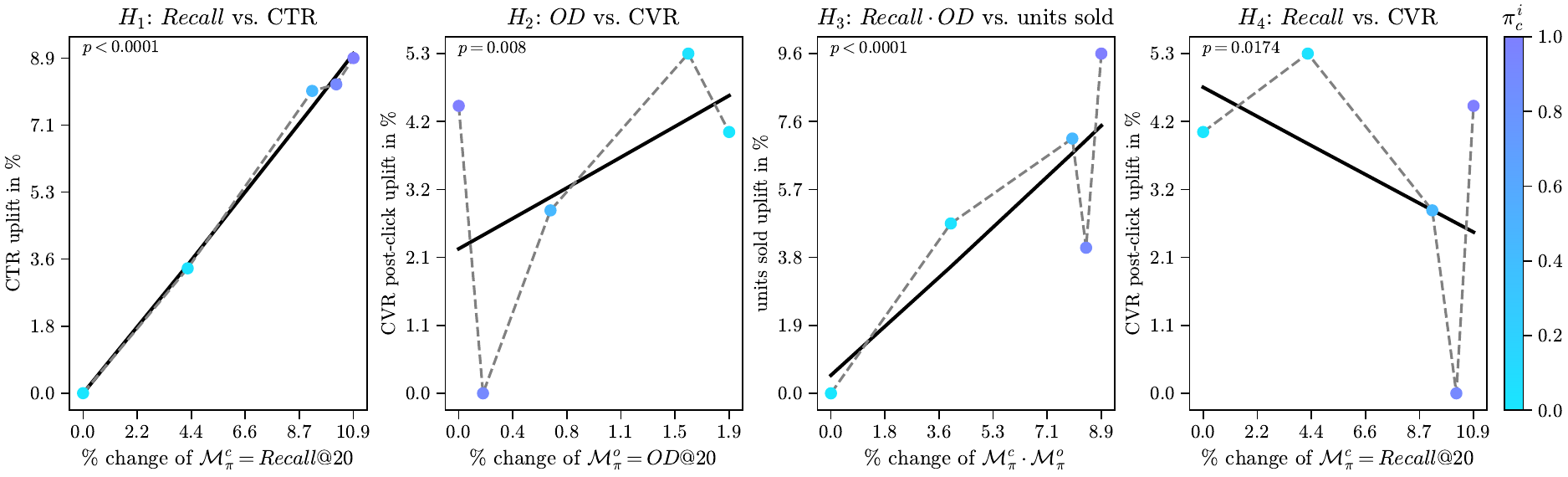}
  \caption{Visualization of the results for hypotheses $H_1$, $H_2$, $H_3$, and $H_4$, shown from left to right, obtained from our experiment. The colored points represent aggregated metrics $\mathcal{M}_{\bm{\pi_i}}$ for each group $i$, each corresponding to a different preference $\bm{\pi_i}=[\pi_c^i, \pi_o^i]$. A black logistic regression line is fitted to the $n$ individual data points across all groups, rather than to the aggregated metrics.
 }
  \label{fig:live_results2}
  \Description[Four line charts]{Four-panel figure showing results of an online experiment linking offline metrics to online KPIs. Each panel plots percent change of an offline metric (x-axis) vs. percent uplift of a KPI (y-axis) across five test groups. Left to right: (1) Recall@20 vs. CTR (x: 0–11\%, y: 0–9\%), shows a strong positive trend; (2) OD@20 vs. CVR (x: 0–2\%, y: 0–5.3\%), shows a moderate positive trend; (3) Recall·OD vs. units sold (x: 0–9\%, y: 0–9.6\%), shows a clear positive relationship; (4) Recall@20 vs. CVR (x: 0–11\%, y: 0–5.3\%), shows a negative relationship. Colored points represent group means; black regression lines are fit to individual data points.}
\end{figure*}

\section{Methods}
\label{strategy}
Let $\mathcal{R}$ be a deep recommender system, and let $\{\mathcal{L}_k(x, y_k)\}_{k=1}^m$ be a set of $m \geq 2$ loss functions, where $x \in \mathbb{R}^N$ and $y_k \in \mathbb{R}^{M_k}$.

\subsection{Pareto Front Approximation}  
Recent advances in Pareto front approximation for deep recommender systems \cite{wilm_pareto_2024} have introduced an efficient approach to learning trade-offs between multiple objectives. A key technique involves sampling a preference vector $\bm{\pi} \sim \text{Dir}(\beta)$ from an $m$-dimensional Dirichlet distribution with parameter $\beta \in \mathbb{R}^m_{> 0}$ during training, incorporating it into the input of $\mathcal{R}$, and scalarizing the objective losses $\mathcal{L}_k$ using $\pi_k$ for $k = 1, \dots, m$. This results in a model $\mathcal{R}(\cdot, \bm{\pi})$ that is conditioned on $\bm{\pi}$ during inference~\cite{dosovitskiy_you_2020,tuan_framework_2024,wilm_pareto_2024}.  

The training objective follows a weighted expectation over $\bm{\pi}$:  
\begin{equation}
 \mathbb{E}_{\bm{\pi}} \mathcal{L}(\cdot, \bm{\pi}, \lambda) =  \mathbb{E}_{\bm{\pi}} \left( \sum_{k = 1}^{m}  \pi_k \mathcal{L}_k\big(\mathcal{R}( \cdot, \bm{\pi}),y_k\big) + \lambda \mathcal{L}_{\text{reg}}(\bm{\pi}) \right),
  \label{eq:paretoloss}
\end{equation}  
where $\mathcal{L}_{\text{reg}}(\bm{\pi})$ is the non-uniformity regulariser \cite{mahapatra_multi-task_2020,wilm_pareto_2024} that ensures a broad Pareto front coverage and $\lambda \ge 0$ is a hyperparameter.

\subsection{Offline Metrics and Online KPIs}
\label{offline-online}
After minimizing Equation~\ref{eq:paretoloss}, offline metrics, denoted as $\mathcal{M}_{\bm{\pi}}$, are evaluated on a holdout test set to analyze $\mathcal{R}(\cdot, \bm{\pi})$ across $g \in \mathbb{N}$ preference vectors $\{\bm{\pi_i}\}_{i=1}^g$. This yields offline metrics $\{\mathcal{M}_{\bm{\pi_i}}\}_{i=1}^{g}$, each corresponding to a specific $\bm{\pi_i}$. To assess online KPIs, traffic is randomly partitioned into $g$ groups. Since $\mathcal{R}(\cdot, \bm{\pi})$ is accessible for any $\bm{\pi}$, a single model can support all preferences. Each group $i$ sends requests to $\mathcal{R}$ with its assigned preference $\bm{\pi_i}$, generating $n_i$ samples $\left(\mathcal{M}_{\bm{\pi_i}}, \mathcal{T}_{\bm{\pi_i}}(j)\right)_{j=1}^{n_i}$, where $\mathcal{T}_{\bm{\pi_i}}(j)$ is the $j$-th online observation for $\bm{\pi_i}$.
The online observations $\mathcal{T}_{\bm{\pi_i}}(j)$ are aggregated to compute the group-level online KPI: $\mathcal{K}_{\bm{\pi_i}} = \text{Agg}_{j=1}^{n_i} \left(\mathcal{T}_{\bm{\pi_i}}(j)\right)$. As $g$ increases, group sizes $n_i$ decrease, leading to higher variance in  $\mathcal{K}_{\bm{\pi_i}}$. To mitigate this, we fit a regression model to the full dataset of metrics and online observations, $\left( \mathcal{M}, \mathcal{T} \right)=\left(\mathcal{M}_{\bm{\pi_i}}, \mathcal{T}_{\bm{\pi_i}}(j)\right)_{i,j=1}^{g,n_i}$, with $n = n_1 + \dots + n_g$ samples. This stabilizes KPI estimates and enables significance testing on the regression parameter of $\mathcal{M}_\pi$, thereby identifying offline metrics predictive of online impact.

\subsection{Strategy Overview}
\label{strategyOverview}
The following steps outline our strategy for identifying which offline metrics reliably predict online KPIs in a real-world system:
\begin{enumerate}
  \item [1.]Train a single model $\mathcal{R}(\cdot,  \bm{\pi})$ to access the full Pareto front.
  \item [2.]Compute the offline metrics $\{\mathcal{M}_{\bm{\pi_i}}\}_{i=1}^{g}$ on the test set.
  \item [3.]Deploy $\mathcal{R}(\cdot, \bm{\pi})$ and split live traffic randomly into $g$ groups.
  \item [4.]Serve each group with its assigned $\bm{\pi_i}$ and obtain $(\mathcal{M},\mathcal{T})$.
  \item [5.]Perform a regression between $\mathcal{M}$ and $\mathcal{T}$ using all $n$ samples, and test whether the relationship is statistically significant.
\end{enumerate}

\subsection{Overcoming Single-Objective Limitations}
In practical applications, a deep recommender system $\mathcal{R}$ may be trained on a single objective, denoted as $\mathcal{L}_1$. Relying solely on a single objective limits the applicability of our approach, which requires at least two distinct objectives. To address this limitation, we introduce an auxiliary \emph{distortion loss} $\mathcal{L}_d$, which acts as a second objective $\mathcal{L}_2$ to artificially induce a trade-off with $\mathcal{L}_1$:
\begin{equation}
  \mathcal{L}_d(x) = \text{CE}(\mathcal{R}(x, \bm{\pi}) \mid \mathbf{1}/c),
  \label{eq:distortionloss}
\end{equation}
where CE denotes the cross-entropy loss, $c$ represents the number of predicted classes/items, and $\mathbf{1}/c = \left[\frac{1}{c}, \dots, \frac{1}{c}\right]$ is a $c$-dimensional constant vector summing to 1.  $\mathcal{L}_d$ pushes the model predictions toward a uniform distribution \cite{mahapatra_multi-task_2020}. This counter-pressure forms the second axis of the Pareto front. It can be seamlessly integrated into existing models, offering a low-overhead solution for practitioners.

\section{Experimental Setup}
 \label{experimentalsetup}
 To validate the proposed strategy from Section \ref{strategyOverview}, we conduct an online experiment on the OTTO e-commerce platform. This experiment aims to assess the strategy's effectiveness in the field of session-based recommender systems by analyzing the relationships between offline metrics and online KPIs. The data structure consists of user sessions, each representing a sequence of user-item interactions $s_{\text{raw}}=[i_{1}^{a_1}, i_{2}^{a_2}, \ldots, i_{T}^{a_T}]$, where $T$ is the session length, and $i_t^{a_t}$ represents the action taken on item $i$ at time $t$ \cite{wilm_pareto_2024}. Actions include clicking or ordering, typically with orders following clicks. Sessions are modeled as $s_{c,o}=[(c_{1}, o_1), (c_{2},o_2), \ldots, (c_{T-1},o_{T-1})]$, where $c_t$ is the clicked item at time $t$, and $o_t$ indicates if the item was ordered in the same session up to time $T$. We address the next-item prediction task \cite{wilm_scaling_2023, hidasi_recurrent_2018, hidasi_session-based_2016} using the sessions $s_{c,o}$, focusing on whether an item is clicked and subsequently ordered. 
 To train the model, we employ two loss functions: the sampled softmax loss, $\mathcal{L}_c$, for predicting item clicks, and the binary cross-entropy loss, $\mathcal{L}_o$, for predicting whether an item is ordered. The neural backbone used is MultiTRON \cite{wilm_pareto_2024}, which employs a session-based Transformer with three layers, a hidden size of 256, a learning rate of $10^{-4}$ with a batch size of 1024, a fixed $\beta = [0.5, 0.5]$, and $\lambda=1$. We utilize a temporal train-test split for training and offline evaluation \cite{hidasi_widespread_2023}.

For \textbf{offline evaluation}, we select $\mathcal{M}_{\bm{\pi}}^c=$ \emph{Recall@20} for the click task. For the order task, we define the \emph{order density} at 20 (\emph{OD@20}) as an offline metric, denoted as:
\begin{equation}
  \mathcal{M}_{\bm{\pi}}^o \coloneq \sum_{i=1}^{C} \mathbf{1}_{\lbrace \text{rank}(c_i) \leq 20, ~o_i=1 \rbrace } \Big/  \sum_{i=1}^{C} \mathbf{1}_{\lbrace \text{rank}(c_i) \leq 20 \rbrace}
\end{equation}
 with $C$ being the number of clicks in the test set. $\mathcal{M}_{\bm{\pi}}^o$ is the empirical probability that a clicked item is ordered, given it is ranked within the top 20 positions.
 The product metric $\mathcal{M}_{\bm{\pi}}^{c,o} \coloneq \mathcal{M}_{\bm{\pi}}^c \cdot \mathcal{M}_{\bm{\pi}}^o$  captures the joint effectiveness of generating clicks and subsequent orders.
 
 For \textbf{online evaluation}, we report three KPIs: $\mathcal{K}_{\bm{\pi}}^c$, the click-through rate (CTR); $\mathcal{K}_{\bm{\pi}}^o$, the post-click conversion rate (CVR); and $\mathcal{K}_{\bm{\pi}}^u$, the total number of units sold. As described in Section \ref{offline-online} and ~\ref{strategyOverview}, we test the following four hypotheses for significance:

\begin{enumerate}
  \item [$H_1$:] \emph{Recall@20} is a positive predictor of CTR.
  \item [$H_2$:] \emph{OD@20} is a positive predictor of CVR.
  \item [$H_3$:] \emph{Recall@20} $\cdot$ \emph{OD@20} is a positive predictor of units sold.
  \item [$H_4$:] \emph{Recall@20} is a negative predictor of CVR.
\end{enumerate}

\section{Results}
We trained the model from Section \ref{experimentalsetup} for 10 epochs on OTTO's private data and conducted the online experiment over the following two weeks in 2024. The traffic was randomly split into five distinct groups ($g=5$), each containing 20\% of the total users. Each group received around 5.3 million impressions, totaling around 26.5 million impressions. Results are reported as percentage changes for both offline metrics and online KPIs, measured relative to the worst-performing group. For each hypothesis presented in Section \ref{experimentalsetup}, we fitted a logistic regression model and tested the significance of the predictor variable $\mathcal{M}_{\bm{\pi}}$ using the Wald test with $\alpha=0.05$. Table \ref{tab:table1} presents the estimated parameters, p-values, and confidence intervals. These results show that \emph{Recall@20} is a significant positive predictor for CTR, \emph{OD@20} for CVR, and \emph{Recall@20} $\cdot$ \emph{OD@20} for units sold. Additionally, \emph{Recall@20} is a significant negative predictor for CVR, consistent with our understanding of the Pareto front. Despite this negative relationship, trading \emph{OD@20} for increased \emph{Recall@20} results in more units sold. Specifically, a 1\% increase in \emph{Recall@20} leads to a 0.9\% increase in CTR, while reducing CVR by only 0.2\%, as shown in Table \ref{tab:table1}. The significant relationships between offline metrics and online KPIs are illustrated in Figure \ref{fig:live_results2}. Notably, CVR exhibits higher variance compared to CTR due to its reliance on low-support events, with absolute differences occurring at the fourth decimal place. This is reflected in the wider confidence intervals presented in Table~\ref{tab:table1}. Units sold also exhibit higher variance due to their implicit dependence on CVR, but benefit from much larger support ($n = 26.5 \cdot 10^6$). These higher variances of CVR and units sold are also apparent in the last three panels of Figure \ref{fig:live_results2}.

\begin{table}
  \centering
  \caption{The results of the experiment for each hypothesis.}
  \setlength{\tabcolsep}{2.75pt}
  \begin{tabular}{ cclc }
    \toprule
    \multirow{1}{*}{\textbf{Hypothesis}}                          & \multicolumn{1}{c}{\textbf{parameter}}  & \multicolumn{1}{c}{\textbf{p-value}}  & \multicolumn{1}{c}{\textbf{[0.025, 0.975]}} \\                                                                                                                                                   
    \midrule
 $H_1$: $\mathcal{M}_{\bm{\pi}}^c$ \, vs. \, $\mathcal{K}_{\bm{\pi}}^c$            & 0.8985                         & $p<10^{-4}$                   & [0.870, 0.927] \\
 $H_2$: ~~~$\mathcal{M}_{\bm{\pi}}^o$ \, vs. \,  $\mathcal{K}_{\bm{\pi}}^o$        & 1.2608                         & 0.0080                        & [0.329, 2.193] \\
 $H_3$: $\mathcal{M}_{\bm{\pi}}^{c,o}$  vs. \,  $\mathcal{K}_{\bm{\pi}}^u$         & 0.7621                         & $p<10^{-4}$                   & [0.557, 0.967] \\
 $H_4$: $\mathcal{M}_{\bm{\pi}}^c$ \, vs. \,  $\mathcal{K}_{\bm{\pi}}^o$           & -0.204                         & 0.0174                        & [-0.372, -0.036] \\
    \bottomrule
  \end{tabular}
  \label{tab:table1}
\end{table}

\section{Conclusion}

This paper presented a pragmatic strategy based on Pareto front approximation techniques for identifying offline metrics that predict online KPIs in real-world recommender systems. The effectiveness of our strategy was validated through a large-scale experiment on the OTTO e-commerce platform, identifying significant relationships between offline metrics and online KPIs. We defined a novel offline metric, \emph{OD@20}, which is a statistically significant positive predictor of the post-click conversion rate. Additionally, the product metric \emph{Recall@20 $\cdot$ OD@20} was proposed as a strong offline proxy for units sold. 
Along the efficient frontier, sacrificing \emph{OD@20} to increase \emph{Recall@20} proved to be a more efficient strategy for driving units sold.
These findings offer valuable insights for industry practitioners, supporting more data-driven decision-making in recommender system optimization. Our results suggest Pareto front approximation techniques as a promising direction for future research aimed at bridging the offline-to-online evaluation gap.

\section{Author Bios}
\textbf{Timo Wilm} is a Lead Applied Scientist at OTTO with ten years of experience, specializing in the design and integration of deep learning models for large-scale recommendation and search systems. He is responsible for translating state-of-the-art research into production-ready solutions within OTTO’s recommendation and search teams, while also contributing to industry research in the field. His work focuses on bridging the gap between academic advancements and industrial applications, ensuring that cutting-edge machine learning techniques drive measurable impact in real-world e-commerce environments.

\textbf{Philipp Normann} is a PhD researcher at TU Wien, working on AI for cybersecurity in the WWTF-funded BREADS project. His research focuses on robust, explainable methods for defending real-world systems. Previously, he spent over seven years as an Applied Scientist at OTTO, developing large-scale AI systems for fraud detection and recommendation. He co-authored several RecSys papers, deployed applied AI solutions across teams and business domains, and led OTTO’s first Kaggle competition with over 2,500 teams, resulting in a widely used public dataset.


\bibliographystyle{ACM-Reference-Format}
\bibliography{paper}

\end{document}